\documentclass[aip,graphicx,amssymb,preprint]{revtex4-1}

\usepackage[utf8]{inputenc}
\usepackage[T1]{fontenc}
\usepackage{graphicx}% Include figure files
\usepackage{mathptmx}
\usepackage{dcolumn}% Align table columns on decimal point
\usepackage{amsmath}
\usepackage{bm}% bold math

\draft % marks overfull lines with a black rule on the right

\begin{document}

%\preprint{}

\title{Improving Molecular Force Fields Across Configurational Space by Combining Supervised and Unsupervised Machine Learning} %Title of paper

\author{Gregory Fonseca}
\affiliation{Department of Physics and Materials Science, University of Luxembourg, L-1511 Luxembourg}
\author{Igor Poltavsky}
\affiliation{Department of Physics and Materials Science, University of Luxembourg, L-1511 Luxembourg}
\author{Valentin Vassilev-Galindo}
\affiliation{Department of Physics and Materials Science, University of Luxembourg, L-1511 Luxembourg}
\author{Alexandre Tkatchenko}
\email{alexandre.tkatchenko@uni.lu}
\affiliation{Department of Physics and Materials Science, University of Luxembourg, L-1511 Luxembourg}

\date{\today}

\begin{abstract}
The training set of atomic configurations is key to the performance of any Machine Learning Force Field (MLFF) and, as such, the training set selection determines the applicability of the MLFF model for predictive molecular simulations. However, most atomistic reference datasets are inhomogeneously distributed across configurational space (CS), thus choosing the training set randomly or according to the probability distribution of the data leads to models whose accuracy is mainly defined by the most common close-to-equilibrium configurations in the reference data. In this work, we combine unsupervised and supervised ML methods to bypass the inherent bias of the data for common configurations, effectively widening the applicability range of MLFF to the fullest capabilities of the dataset. To achieve this goal, we first cluster the CS into subregions similar in terms of geometry and energetics. We iteratively test a given MLFF performance on each subregion and fill the training set of the model with the representatives of the most inaccurate parts of the CS. The proposed approach has been applied to a set of small organic molecules and alanine tetrapeptide, demonstrating an up to two-fold decrease in the root mean squared errors for force predictions of these molecules. This result holds for both kernel-based methods (sGDML and GAP/SOAP models) and deep neural networks (SchNet model). For the latter, the developed approach simultaneously improves both energy and forces, bypassing the compromise to be made when employing mixed energy/force loss functions.
\end{abstract}

\pacs{}% insert suggested PACS numbers in braces on next line

\maketitle %\maketitle must follow title, authors, abstract and \pacs

%%%%% INTRODUCTION
\section{Introduction} \label{Introduction}

With the enormous rise in computational power and molecular simulation methods in the last decades, atomistic modeling is increasingly becoming the method of choice~\cite{Hollingsworth2018, Phillips2005, HANSSON2002190, ABRAHAM201519, rapaport2004art, Plattner2015, Krylova2020, Wu2019, Wolf2019}. Applications range from the study and prevention of corrosion~\cite{Kallikragas2018, DorMohammadi2018, Obot2019} to protein folding~\cite{Best2012}, unfolding~\cite{Xiao2018}, and self-assembly~\cite{Meneksedag-Erol2019}. 

Efficient molecular modelling can be done using empirical force-fields (FF): computing interactions between atoms and molecules using predefined functional forms. Unfortunately, when it comes to accuracy, such simulations leave a lot to be desired; they are, for example, not able to compute any chemical changes or capture many-body effects in a reliable way. When both accuracy and efficiency are a concern, machine learning force fields (MLFF) are becoming the method of choice.

In contrast to empirical FF, ML models can potentially reproduce any functional form of interatomic and intermolecular interactions, leading to reliable descriptions of potential energy surfaces (PES) of arbitrary complexity. Many successes have been found in this domain in the recent years, with a multitude of methods being able to predict the behaviour of small to medium sized molecules and more~%\cite{Noe2020, sgdml, Schutt2017, Montavon2013, Behler2016, Schutt2017a, Hansen2015, Mardt2018, Imbalzano2018, PhysRevLett.108.058301, inbook, NIPS2017_7232, C7SC04934J, doi:10.1021/acs.jctc.7b00577, doi:10.1021/acs.jpcb.7b09636, doi:10.1063/1.4712397, PhysRevB.95.214302, doi:10.1021/acs.jpcc.8b09917}. 
\cite{Noe2020, sgdml, Schutt2017, Montavon2013, Behler2016, Schutt2017a, Hansen2015, Mardt2018}
These methods were used to calculate the stability of molecules with chemical accuracy~\cite{Bartok2017}, predict the formation energy of crystals at the level of density functional theory~\cite{Faber2016}, or even reconstruct phase diagrams~\cite{Artrith2018}, to name a few examples. 

Despite those achievements, the data-driven nature of ML has its downsides: the quality of the ML results are at the mercy of the availability of ``good'' initial data. Collecting such data and choosing ``good'' training points is a nontrivial problem requiring a deep understanding of the nature of the data, which relies on human intuition. This puts into question the unbiased nature of the ML approaches, eliminating one of their main advantages over the human-designed FFs. For instance, for applications in molecular dynamics (MD) simulations, the training data are generally parts of molecular trajectories extracted from a reference \textit{ab initio} simulation with the desired level of accuracy. ML models are then frequently trained to have the best overall prediction across the entire dataset. This however skews the ML models toward more common (close-to-equilibrium) molecular configurations, as poorly predicted but rare (out-of-equilibrium) configurations hardly impact overall statistics. Hence, the usage of such ML models is unreliable and unpredictable in long MD simulations where out-of-equilibrium configurations are significantly more important for constructing the partition function. This will always be the case when the data distributions in the target simulations are different from those in the reference datasets. Examples include: studying nuclear quantum effects (such as proton transport) using the MLFF trained on classical MD trajectories, simulating phase transitions based on the information collected only in stable phases, computing reaction rates with ML models trained on meta-dynamics, etc. In all these cases, minimizing the prediction error on the reference dataset does not guarantee good prediction quality across all important configurations, making the results of the simulations questionable.

In this work, we address the issue outlined above by ``flattening the error'' of ML models: i.e. we ensure that the predictive accuracy of the MLFF is equally reliable for out-of-equilibrium structures or rare events as for common configurations, thus enhancing the stability of the model regardless of its use case. To accomplish this, we propose a novel method to optimise the training of ML models, leading to unbiased molecular FFs with almost constant accuracy across the entire reference dataset. This method is equally applicable to any ML model and is available in our free open-source MLFF package~\cite{github}. We showcase its application on small organic molecules (uracil, salicylic acid, ethanol, toluene) as well as a larger molecule (alanine tetrapeptide) using GAP~\cite{GAP} models with the SOAP~\cite{SOAP} descriptor and sGDML~\cite{sgdml} as a representative kernel-based approach and SchNet~\cite{Schutt2017} as a representative neural-network-based approach. Comparing our improved models to default models of equal training set size reveals an error reduction on rare/out-of-equilibrium configurations by a factor of up to 2 for a negligible sacrifice in mean error. 

While most standard approaches involve calculating an average error across an entire dataset, we are able to compute the prediction error of a model for different regions of the reference data (configurational space for our examples), leading to a detailed view on the domain of applicability of the model. The presented approach can also be applied as an outlier detection method, effectively finding rare processes or out-of-equilibrium mechanisms inside of a dataset in an automated way. As an example, the developed approach enabled us to reveal the fingerprints of the proton transfer mechanism between the hydroxyl and the carboxylic group in the salicylic acid molecule database~\cite{sgdml} generated using classical MD simulations. This process is represented only by a few hundreds of configurations within more than 320k molecular geometries. Nevertheless, the developed approach is sensitive enough to separate this subset of geometries into an individual cluster.

The structure of the article is the following. In the \textit{Theory} section we explain the developed methodology for outlier detection and improved training technique. The \textit{Practical Application} section contains the best practice example, where we explain in detail how to use the proposed method for the outlier detection and improved training on the example of the salicylic acid molecule. In the \textit{Results} section we apply our method to reconstruct the FF of small organic molecules and alanine tetrapeptide, which serves as a representative case for how the method performs on larger molecules. The \textit{Conclusions} section presents a summary and an outlook.

%%%%% THEORY
\section{Theory} \label{Theory}

\begin{figure*}[h!]
    \centering
    \includegraphics[width=0.9\textwidth]{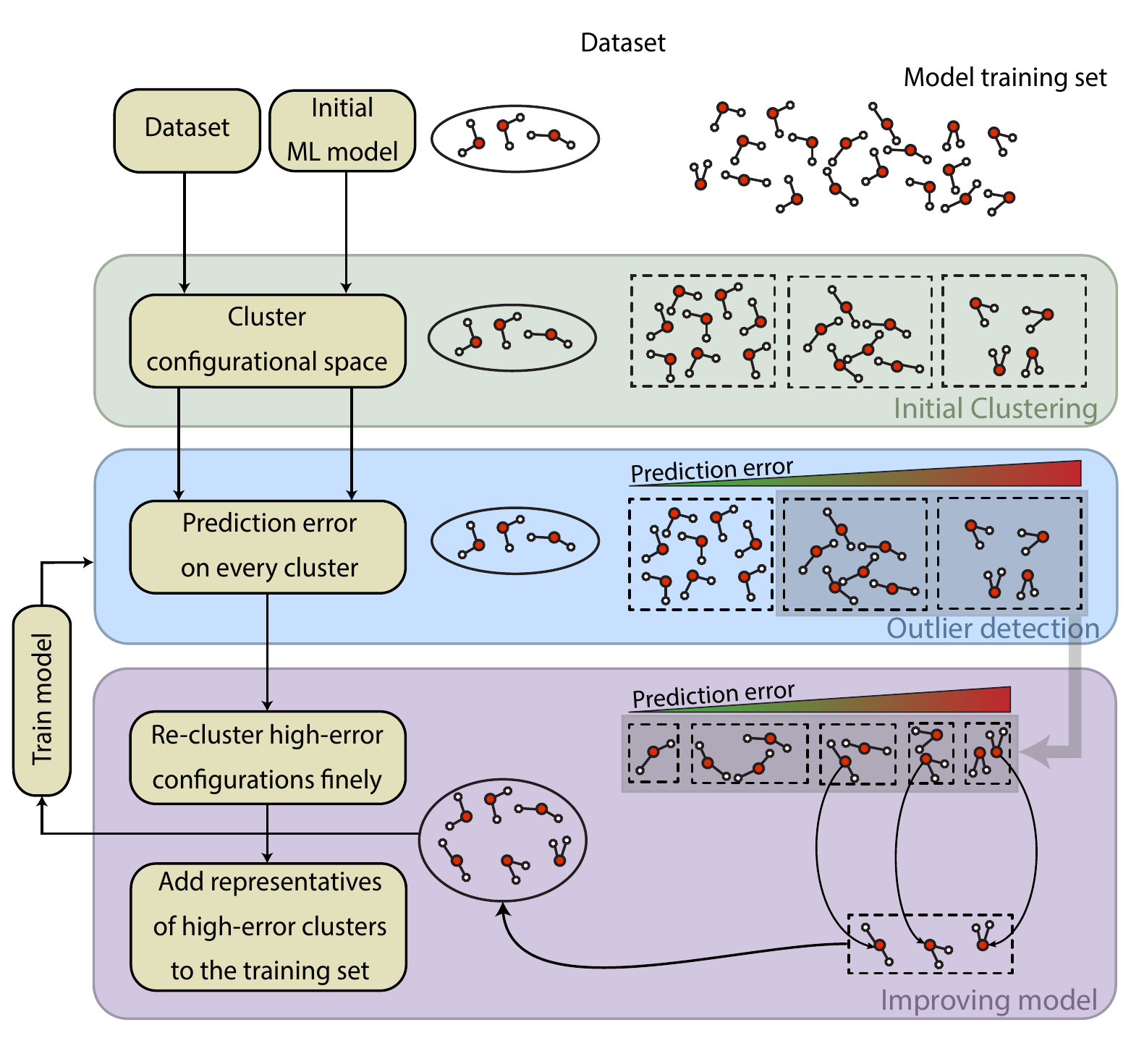}
    \caption{Overview of the improved learning method. A dataset is clustered into subsets and the error of an initial ML model is assessed on each individual cluster. High-error clusters are re-clustered finely and representative configurations are extracted from each and added to the training set. This is repeated until a given number of training points is reached.}
    \label{fig:roadmap}
\end{figure*}

The method described in this section allows for an in-depth error analysis of a ML model, outlier detection, as well as an improved training technique resulting in ``equally'' reliable predictions for all parts of the entire reference dataset. We apply the developed approach for constructing accurate MLFF for molecules consisting of a few tens of atoms, but it can be generalized for any regression problem. An overview of the methodology can be found in Figure~\ref{fig:roadmap}. 

The method can be subdivided into three main steps. In the ``Initial Clustering" step, we split molecular configurations into groups based on similarities in geometric and energetic properties using a combination of clustering techniques. Namely, we employ agglomerative clustering~\cite{Ward} to group up configurations of similar geometries and further split the groups in different energy brackets using KMeans~\cite{Sculley2010}.

We then apply a given ML model to each individual cluster and compute the respective mean prediction error, as illustrated in the ``Outlier detection" step. Clusters with high prediction errors represent regions of configurational space (CS) the model is ill-adapted to, restricting its range of applicability. This poor prediction can arise from two main sources: significant differences in physical/chemical properties compared to common configurations, and/or poor representation for specific regions of CS in the training set. 

We address the latter in the ``Improving model" step, where the combination of all poorly predicted areas is considered and subdivided into a larger amount of clusters, providing a fine grid of the problematic regions. The numerous clusters allow us to filter out well-predicted configurations that the initial clustering had previously misrepresented, as well as find problematic clusters on a finer scale. Extracting representative geometries (largest error, cluster centroid, random,...) from poorly predicted fine clusters and adding them to the initial training set improves the model's performance on the problematic regions of CS. Repeating the described procedure with re-trained models results in a final model with an optimised training set of a given size, capable of producing comparable errors across all reference data. 

% split molecular configurations (clustering)
In order to subdivide tens of thousands of unlabeled data points into just a handful of broad initial clusters, similarities between molecular configurations were defined based on the descriptors of CS used for training the ML models. In this paper, pairwise atomic distances is the descriptor of choice. Differences between configurations were defined using the Euclidean distance in the descriptor space. An agglomerative approach was chosen to cluster the dataset into configurations with similar geometries, as the algorithm avoids merging rare but geometrically unique configurations with large groups of common ones.

With $\vec{x_i^a}$ the Euclidean position of atom $i\in[1,N]$ of data point $a\in[1,M]$, the descriptor $\vec{z^a}$ is given by:
\begin{align}
    \vec{z^a}&= [...,z^a_{i,j},...], \; j<i \\
    z^a_{i,j}&=||\vec{x_i^a}-\vec{x_j^a}||_2
\end{align}

with $||\cdot||_2$ a simple Euclidean distance. Distances in the descriptor space are then:
\begin{align}
    d(\vec{z^a},\vec{z^b})= ||\vec{z^a}-\vec{z^b}||_2
\end{align}

Since Euclidean distances are not a natural metric of our chosen descriptor space, clusters produced this way often contained large variations in potential energy. To avoid this problem, a further distinction between different energy levels was done using a KMeans method. The combination of both clustering techniques helped distinguish between possible degenerate states as well as geometrically `similar' configurations with significant energy differences. 

After successfully splitting the dataset both by geometries and energies, an initial ML model was applied to every individual cluster and the average error on all configurations therein was computed between the predicted and actual forces. The root mean squared error (RMSE) was chosen as a way to emphasize large differences. Ordering the clusters by their average prediction error led to a simple way to identify outliers for the given dataset and model. Poor predictions on specific clusters were commonly caused by the training set containing too few examples from the relevant region of CS. Often, this arose as a simple consequence of a non-optimal training set choice: out-of-equilibrium geometries are naturally rarer and thus less represented in datasets born from physical simulations. As such, a random choice of training points --- even if according to some statistical distribution --- is very unlikely to contain those important out-of-equilibrium points. In other cases, clusters contained configurations whose physiochemical properties deviate from the rest of the dataset. In such cases, even small changes in geometry can lead to large differences in forces, hence the need to include a sizable contribution of outlying configurations to the training set for accurate predictions.

The outlier detection described above also enabled an improved method to choose the training set. To this end, poorly predicted initial clusters were recombined and re-clustered more finely by applying the agglomerative approach as before but with a larger amount of clusters. This increased the resolution in which problematic regions of CS were identified, allowing for (a) filtering out of well-predicted configurations, previously buried in overly broad clusters, and (b) a finer distinction between all sub-regions of CS that include the configurations problematic for our initial model. Systematically adding data points from the worst predicted fine clusters to the model's training set ensured that all subregions of CS were sufficiently represented. 

Several methods were explored to choose which data points to add from the fine clusters to the training set. Selecting random points from the clusters already lead to improvements, but to a lesser extent than centroids (in the descriptor space) or points with the highest prediction error within their cluster. Both of the latter methods performed similarly, however, as previous steps already required the computation of prediction errors for every point, the highest-prediction error criterion proved to be more efficient and is the default method for this paper. 

An alternative to the above scheme could be considered: to skip the initial clustering and simply continue the process using all configurations whose prediction error exceeds some minimum value. However, this would require calculating the prediction for every single data point, whereas our method allows calculating only a subset of each cluster as a representative error for the whole, saving on valuable computational cost. As the datasets in this work are of limited size, the latter was not necessary, but will become important when scaling to larger systems.
Furthermore, many well-predicted clusters still contain singular configurations associated with high errors. As opposed to poorly predicted configurations inside poorly predicted geometry clusters, the former are victims to the limitations of the ML model rather than that of the training set. Including those in the training set in an attempt to improve their prediction comes at a significant cost in accuracy in their otherwise well-predicted cluster. Including entire clusters rather than singular points of high error anchors the aforementioned exceptions to their respective low-error cluster, thus giving room to the algorithm to favour configurations that are representative of entire poorly-predicted subregions of the CS.

One could also think of skipping the initial clustering step and immediately proceed to creating fine clusters from which to extract new training points. However, our clustering method of choice---agglomerative clustering---has a time complexity of $O(n^3)$ and memory requirement of $O(n^2)$, making it inadequate for handling large datasets in one go; instead in a first step, a subset of the data is chosen and clustered. The remaining (unclustered) data points are then iteratively added to an existing cluster based on smallest average distance, mimicking agglomerative clustering while bypassing computational limitations. This approach reduces the quality of the clustering scheme, but is still able to exclude well-predicted regions of CS in broad strokes. The combination of all remaining clusters represent a subset much smaller than the original dataset, thus most, if not all of the remaining data points can now be clustered in a single agglomerative step, leading to fine clusters of higher quality. 

Thanks to the fine grid of problematic configurations provided by our clustering algorithm, new data points could be added to the training set such as to address the model's poor predictions in targeted way. The complete training set for given data was created in an iterative manner, successively computing prediction errors, targeting problematic configurations to add to the training set, and re-training ML models. In the end, resulting models were trained on all the necessary data points to produce comparable prediction errors across all of CS within a dataset. This extends the MLFF application range beyond near-equilibrium simulations, providing reliable results even for out-of-equilibrium computations such as finding reaction rates or transition pathways.

%%%%% BEST PRACTICES
\section{Practical application of clustering algorithm to salicylic acid ML force field} \label{BestPractices}

In this section we describe in detail each step of the outlier detection and improved training process on the example of the salicylic acid molecule~\cite{sgdml}. All the results shown within this article were obtained using exactly the same procedure and settings as explained here unless specified otherwise. 

As a first step, the atomic positions of each reference configuration are converted to the more appropriate pairwise distance descriptor. This descriptor is used to split the dataset into 10 clusters through agglomerative clustering with Ward~\cite{Ward} linkage and Euclidean metric. Then, an additional clustering step is performed on each individual cluster using the KMeans algorithm with Euclidean metric and kmeans++ initialization~\cite{kmeans++}. This step splits each previous cluster into 5 for a total of 50 clusters. 

\begin{figure*}[h!]
    \centering
    \includegraphics[width=0.49\textwidth]{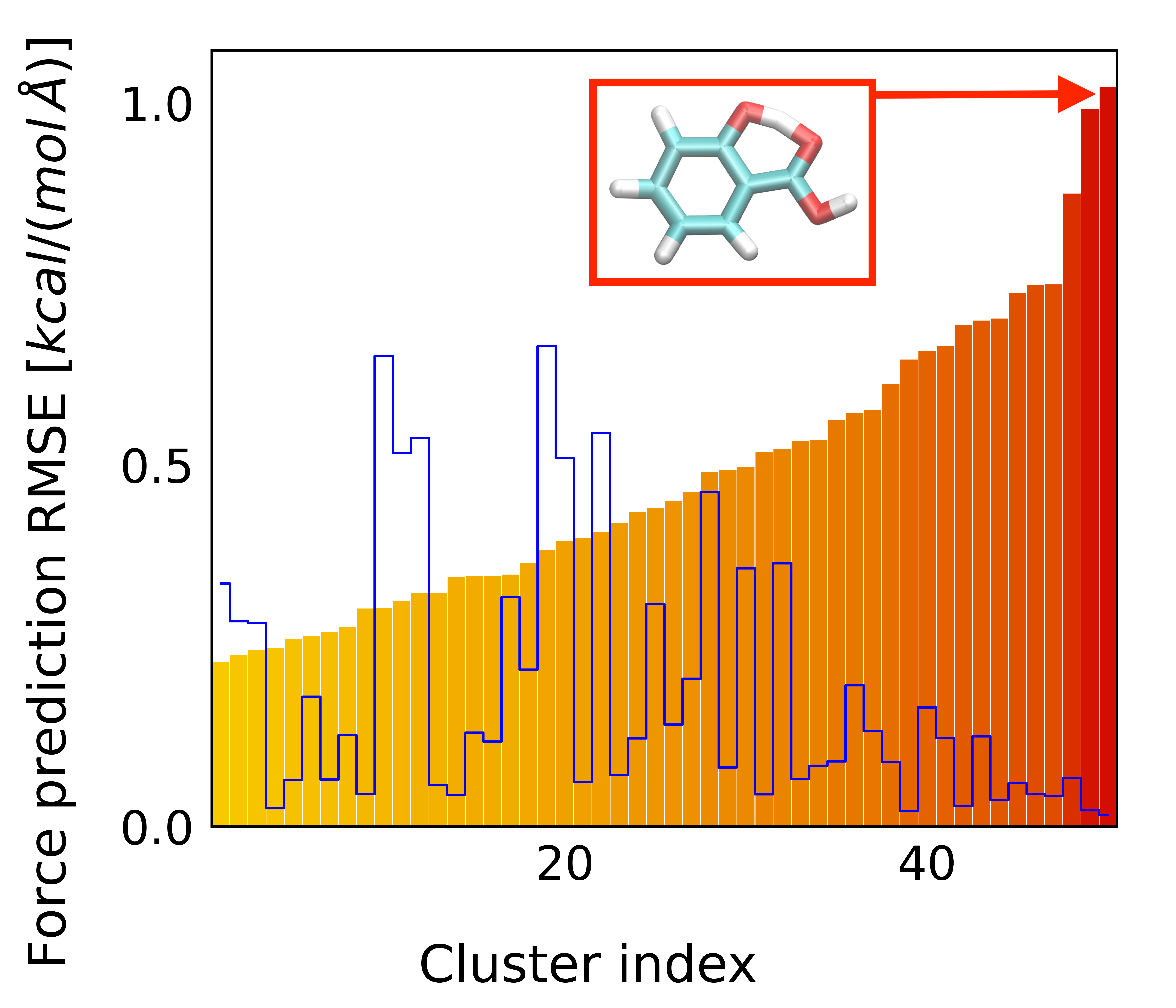}
    \caption{Force prediction root mean squared error (bars) on all 50 clusters (x-axis) of the salicylic acid dataset, ordered by ascending error. Relative population of each cluster is also indicated (solid blue line, arbitrary units). A representative structure of the highest-error cluster is shown (red box).}
    \label{fig:CE_SAL}
\end{figure*}

For purposes of outlier detection, an sGDML model is trained on the salicylic acid database~\cite{sgdml} with 1000 training points using the default training scheme implemented within the sGDML package. This model is subsequently used to predict forces for all 320k configurations of the reference dataset. Then, force prediction errors are computed for each individual cluster, and clusters are rearranged based on the cluster RMSE (see Figure~\ref{fig:CE_SAL}). This outlier detection can be done automatically for sGDML using our MLFF software~\cite{github} with the following call on default settings:

\begin{center}
\begin{verbatim}
python run.py cluster_error -d <dataset_file> -i <model_file>
\end{verbatim}
\end{center}

The above will provide the user with a graph similar to Figure~\ref{fig:CE_SAL} as well as the indices of every cluster and the prediction errors on each one respectively. It is worth noting that each cluster corresponds to a qualitatively different set of configurations. Hence the proposed scheme detects poorly predicted regions of CS for a given model. An example geometry from the worst-predicted cluster is shown in the figure below: this configuration has a clear fingerprint of shared hydrogen between the carboxylic and hydroxyl groups. This process is a rare event in the reference database obtained by employing classical MD simulations and can be easily missed by visualization of the trajectory or other human analyses. In contrast, the proposed clustering approach can easily separate such nontrivial configurations (a few hundred) from the overwhelming number (above 300 thousand) of simple fluctuations around the equilibrium geometry.

In order to create improved models, we instead start with a smaller sGDML model trained on only 200 configurations using the default training scheme. The same error analysis is performed, giving us a rough idea of which part of CS the current model is struggling with. Every cluster whose error exceeds a factor of the overall error (factor of $1.1$ here) is merged and re-clustered to distinguish between finer subparts of CS. The fine clusterisation created a total of 200 clusters based solely on pairwise atomic distances, using the same agglomerative approach as for the initial clusterisation. Finally, one representative configuration is extracted from half the fine clusters (prioritising high-error clusters). The extracted configuration corresponds to the highest error within the respective cluster. Overall, 100 points are extracted to be added to the training set before the sGDML model is re-trained using the new combined training set (containing 300 configurations in total). The model's errors are then re-assessed on the same initial broad clusters, after which new fine clusters are created and a new subset of 100 training points is extracted. This process is repeated 8 times in total, resulting in an optimized training set of 1000 reference molecular configurations and the corresponding improved sGDML model. Using our software, one can simply execute the following command:
\begin{center}
\begin{verbatim}
python run.py train -d <dataset_file> -n 8 -i 200 -s 100
\end{verbatim}
\end{center}

\begin{figure*}[h!]
    \centering
    \includegraphics[width=1\textwidth]{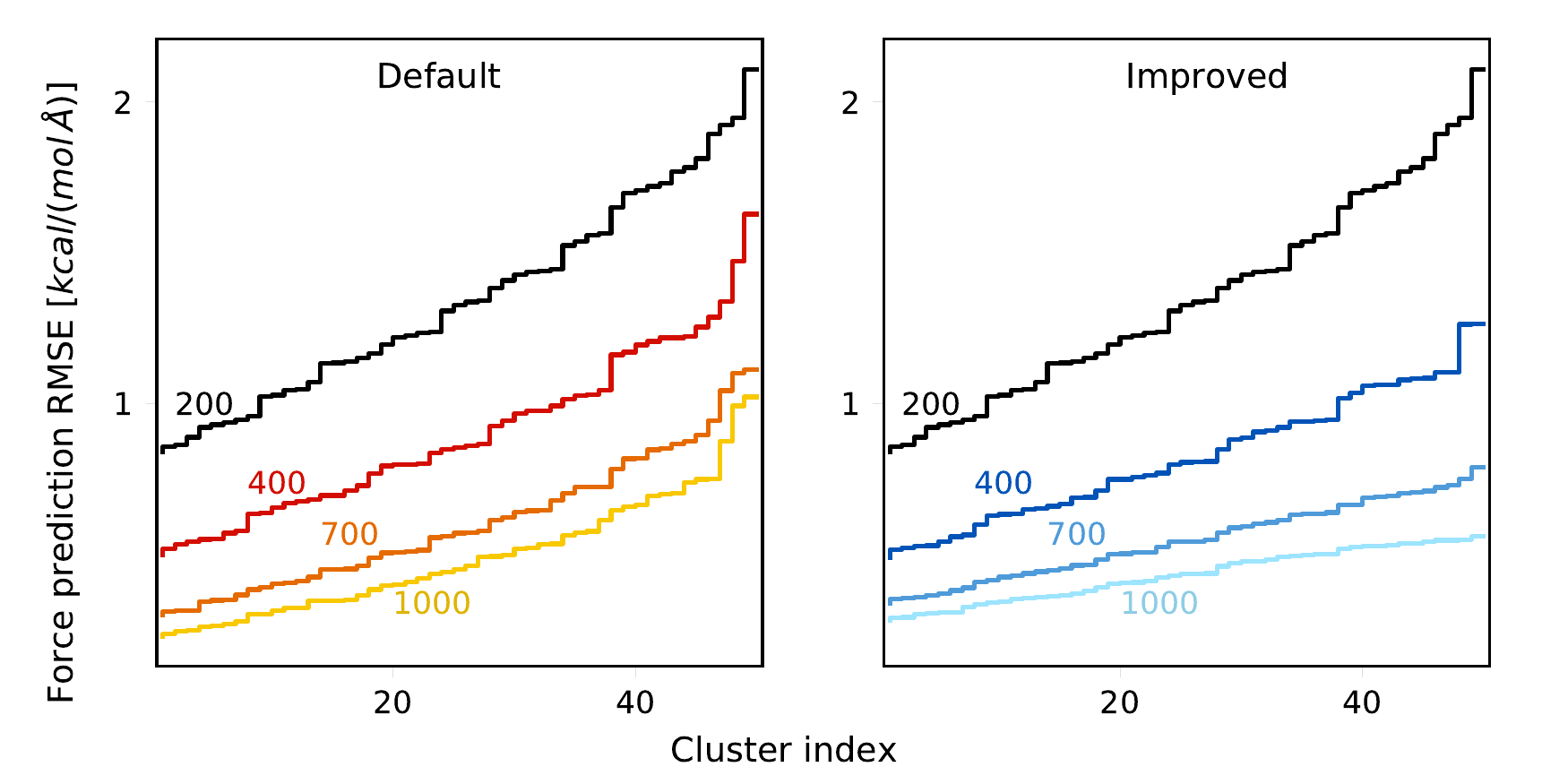}
    \caption{Force prediction root mean squared error (solid lines) on all 50 clusters (x-axis) of the salicylic acid dataset, ordered by ascending error. Different colours correspond to varying sizes of training set, using the default sGDML training method (left) and the improved method (right).
    }
    \label{fig:SAL_TRAIN}
\end{figure*}

The above command will provide the improved model with 1000 training points as per the procedure described above, as well as the indices of the broad clusters. The results are shown in Figure~\ref{fig:SAL_TRAIN} for the initial, final, and one intermediate step.
For comparison we add the results of the default training scheme as implemented in the sGDML package for models of equal size. One can see a noticeable flattening of the error curve with every iteration step of the improved training method.  In contrast, the increase in the number of training points for the default sGDML model leads mainly to an overall decrease in prediction errors, keeping entire parts of the CS poorly predicted.

%%%%% RESULTS
\section{Results} \label{Results}
The developed methodology was used to perform a detailed error analysis of three state-of-the-art MLFF models, namely sGDML~\cite{sgdml}, GAP~\cite{GAP} using the SOAP~\cite{SOAP} descriptor and SchNet~\cite{Schutt2017}. The reference datasets used are of ethanol, salicylic acid and uracil~\cite{gdml}. From Figure~\ref{fig:CE_combined} (note the different scales of the ordinate axes), it is clear that these different models show varying performance on the molecular datasets, however all of them are consistently inaccurate for out-of-equilibrium geometries. We show that by employing default training techniques, the prediction error on some physically relevant configurations can exceed the overall root mean squared error (RMSE) by up to a factor of 3. On the flip side, our improved training method alleviates the problem by creating models with significantly flattened errors across all configurations. This was applied to the previously mentioned datasets as well as toluene~\cite{gdml} and an alanine tetrapeptide dataset. 

In many cases, the RMSE of initially poorly predicted configuration clusters is reduced twofold, while well-predicted clusters suffer a marginal increase in RMSE, largely within the error margin of the original dataset. Furthermore for most molecule/model combinations, our improved learning methods result in an overall RMSE decrease despite the focus of the method on rarer configurations. 

\subsection{Outlier detection}\label{OutlierDetection}

\begin{figure*}[ht!]
    \centering
    \includegraphics[width=.9\textwidth]{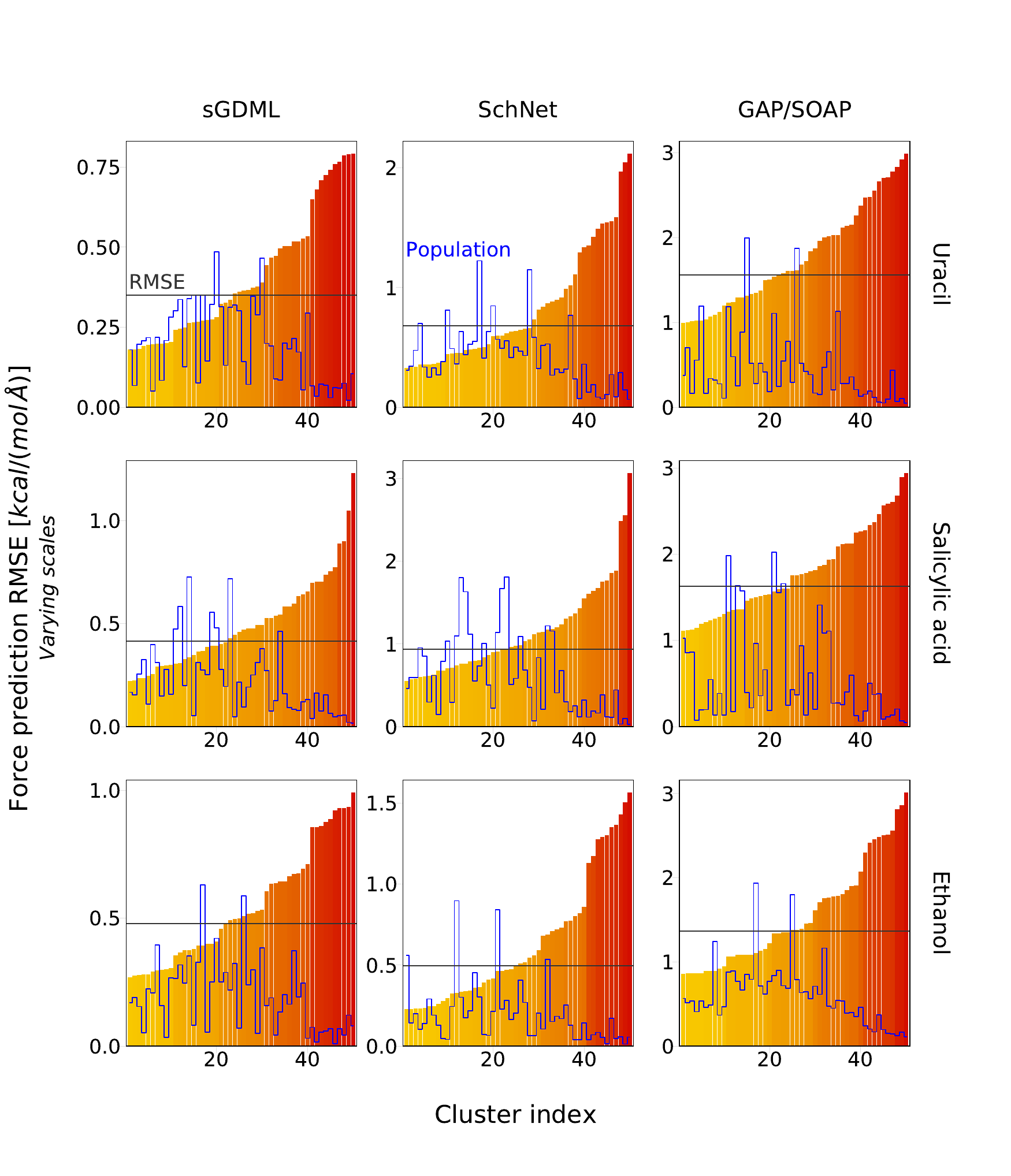}
    \caption{Force prediction RMSE for sGDML, SchNet and GAP/SOAP (with 12 radial and 6 angular functions) models on the same ethanol, uracil and salicylic acid datasets (y-axis, scale adapted for each model for better visibility), split into 50 clusters of similar configurations (x-axis) ordered by ascending error. RMSE (bars) is given on a per-cluster basis in contrast to the RMSE over the entire dataset (solid horizontal black line). Relative cluster populations are also indicated (solid blue line, arbitrary units). }
    \label{fig:CE_combined}
\end{figure*}

Before improving the models, it is necessary to find and show the underlying problems in the default training methods. For this, we used our outlier detection methods on multiple datasets for different molecules. The MLFF models of choice were sGDML, SchNet and GAP/SOAP: we applied each model to the same molecular datasets. We cluster a dataset of salicylic acid, uracil and ethanol into 50 different regions of CS and compute the mean squared force prediction error for every cluster. The results are plotted in Figure~\ref{fig:CE_combined}. A very large disparity between the errors in clusters and the mean squared error can be observed, with some clusters presenting an error 3 times higher than the mean. The difference between them and the cluster of lowest error is of course even higher. It is clear that in these cases, a single MSE is a very bad metric to quantify how good the ML model works for out-of-equilibrium geometries. This is in direct contradiction with the idea of the MLFF being comparable to the underlying \textit{ab initio} method, as entire regions of CS present an accuracy significantly worse than that of the reference calculations.

There are two possible reasons for the observations above. The trivial one is that the poorly predicted regions contain large fluctuations of molecular geometries, which are not well represented in the training set. This lack of information then renders the ML models unable to learn them. Many applications mainly deal with close-to-equilibrium molecular configurations where the PES of the molecule is constrained within a given region of CS. There, higher prediction errors out-of-equilibrium might not significantly impact the results of simulations. However, for studying molecular stability, configurational changes or chemical reactions to name a few, large prediction errors on such fluctuations may affect the final results greatly. 

The second reason is non-trivial and can have a significant impact on the reliability of the MLFF. The poorly predicted areas of the CS can represent physics or chemistry missing in the majority of the configurations in the reference dataset. This is showcased in our previous example of salicylic acid, where the cluster with the most significant error corresponds to a shared proton between the carboxylic and hydroxyl group (for the details see the \textit{Practical Application} section). An accurate simulation of this process would require proper account of nuclear quantum effects, hence the corresponding configurations are a negligible minority (a few hundred) in the salicylic acid dataset containing over 320k molecular geometries of a classical MD run. Even if corresponding reference data were added to the MD dataset, those would mainly be ignored within the standard training schemes due to their relatively high energies. All in all, this points to default MLFF being inapplicable for studying the proton sharing effect for our given dataset. In contrast, the developed method is designed to alleviate this problem by widening the model's applicability range to the fullest capability of the dataset.

\subsection{Improved models}\label{Improved_models}
\begin{figure*}
    \centering
    \includegraphics[width=.90\textwidth]{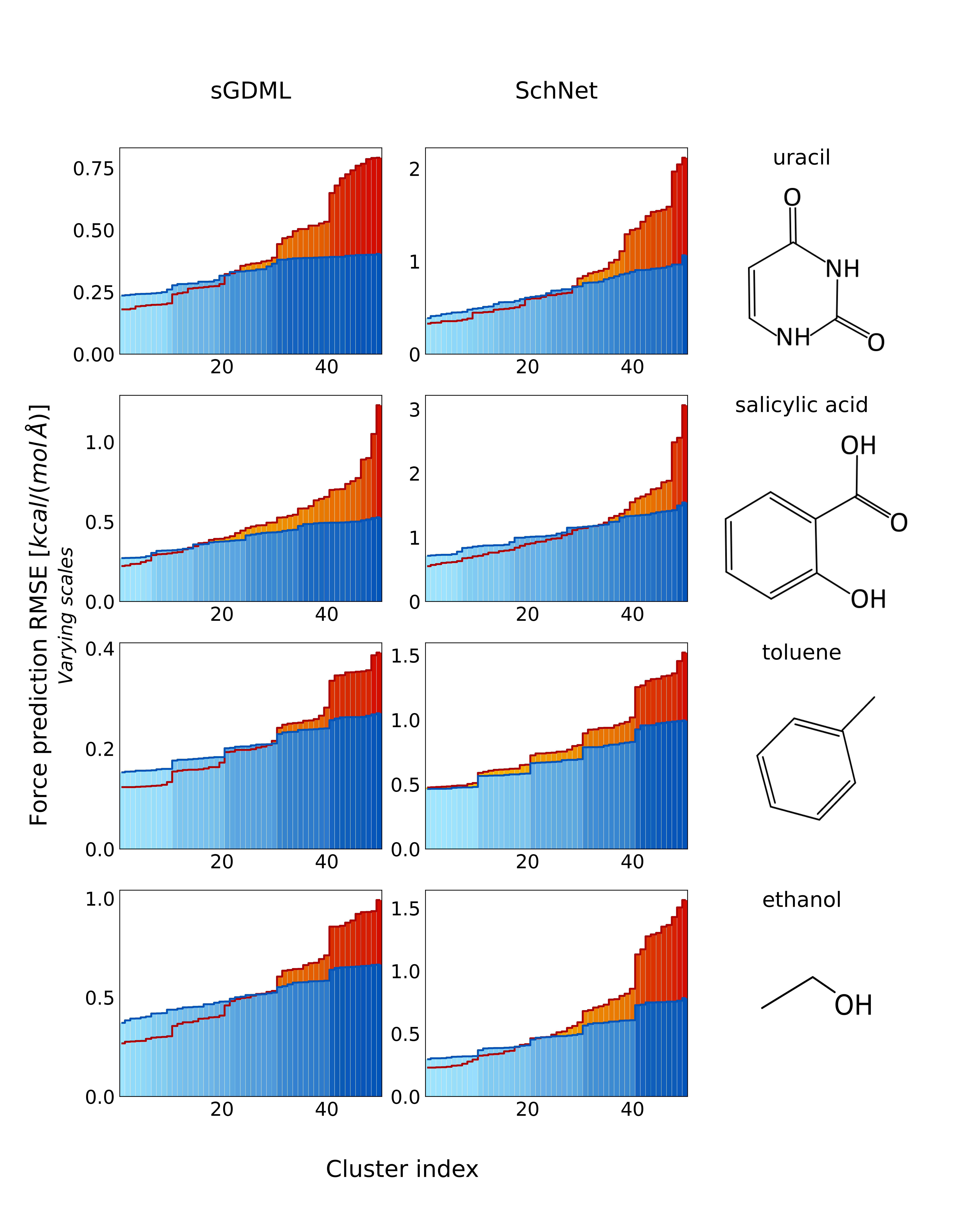}
    \caption{Force prediction RMSE for sGDML and SchNet default models compared to the improved models (orange/blue bars, y-axis scale adapted for each model for better visibility). RMSE was computed on a per-cluster basis on ethanol, uracil and salicylic acid datasets, split into 50 clusters of similar configurations (x-axis) ordered by ascending error.}
    \label{fig:IL_combined}
\end{figure*}

To resolve the problem with nonuniform prediction accuracy, we applied the improved training techniques developed in this work using both sGDML and SchNet as our FF models once again. The molecules explored include all the datasets from the previous subsection as well as toluene. First, we performed the outlier detection by computing the root mean squared force prediction error on 50 clusters for an initial model with 200 training points. After that, 100 training points were added at every step for a total of 8 steps, resulting in models of a total of 1000 training points each. All the details of the improved training procedure can be found in the \textit{Practical Application} section. The comparison between the default and improved models of the same size is shown in Figure~\ref{fig:IL_combined}. In contrast to the default models, the improved versions present a more constant accuracy, with most models reaching a maximal cluster error of less than half that of their default counterpart. The results shown in Figure~\ref{fig:IL_combined} represent the quality of MD simulations performed with the default and the improved models with respect to the reference method. Since forces are the variables entering the equations of motion, their errors are directly related to the deviations between the reference and ML trajectories (more so than the energies). Of course, computing properties that are mainly defined by the most common configurations in the reference dataset---such as average energies at reasonably low temperatures---both types of models would lead to nearly identical results. On the other hand, processes involving broad parts of the PES or regions underrepresented in the reference dataset will be much better described using the proposed improved ML models.

The goal of the improved models is to present a more stable prediction error across all of configurational space. They do so by explicitly including more out-of-equilibrium/rare configurations in their training set, at the expense of the more common/in-equilibrium configurations in the dataset. Despite this, the overall RMSE across the entire dataset does not change significantly, and even sees some decrease for many of the molecules shown below (see Table~\ref{table:1}). This further highlights the importance and usefulness of choosing the training set in a careful and meticulous way beyond just RMSE. 

\begin{table}[h!]
\centering
\caption{Overall RMSE for sGDML and SchNet models, comparing default and improved versions. All numbers are given in $kcal/\left(mol\,\AA\right)$}
\begin{ruledtabular}
\begin{tabular}{ccccc}
molecule & def. sGDML & imp. sGDML & def. SchNet & imp. SchNet \\
\hline 
uracil & 0.38 & 0.32 & 0.77 & 0.65 \\
salicylic acid & 0.44 & 0.39 & 0.99 & 1.03 \\
toluene & 0.21 & 0.20 & 0.78 & 0.67 \\
ethanol & 0.51 & 0.50 & 0.57 & 0.47 \\
\end{tabular}
\end{ruledtabular}
\label{table:1}
\end{table}

\subsection{Application to larger molecule}

So far, we applied the developed methods only on rather small molecules, demonstrating significant improvements in the resulting MLFFs. In this subsection, we extend the applications to noticeably larger molecules, using as an example an alanine tetrapeptide (AcAla3NHMe). This peptide is large enough to exhibit several incipient secondary structure motifs akin to biological peptides and proteins. It is important to note that, especially for larger molecules, our training method can only lead to improvements if the base model has acceptable accuracy in the first place, hence we use SchNet as our model in this subsection, since SchNet can be easily employed with much larger datasets than kernel-based methods (sGDML or GAP).

\begin{figure*}[h!]
    \centering
    \includegraphics[width=1\textwidth]{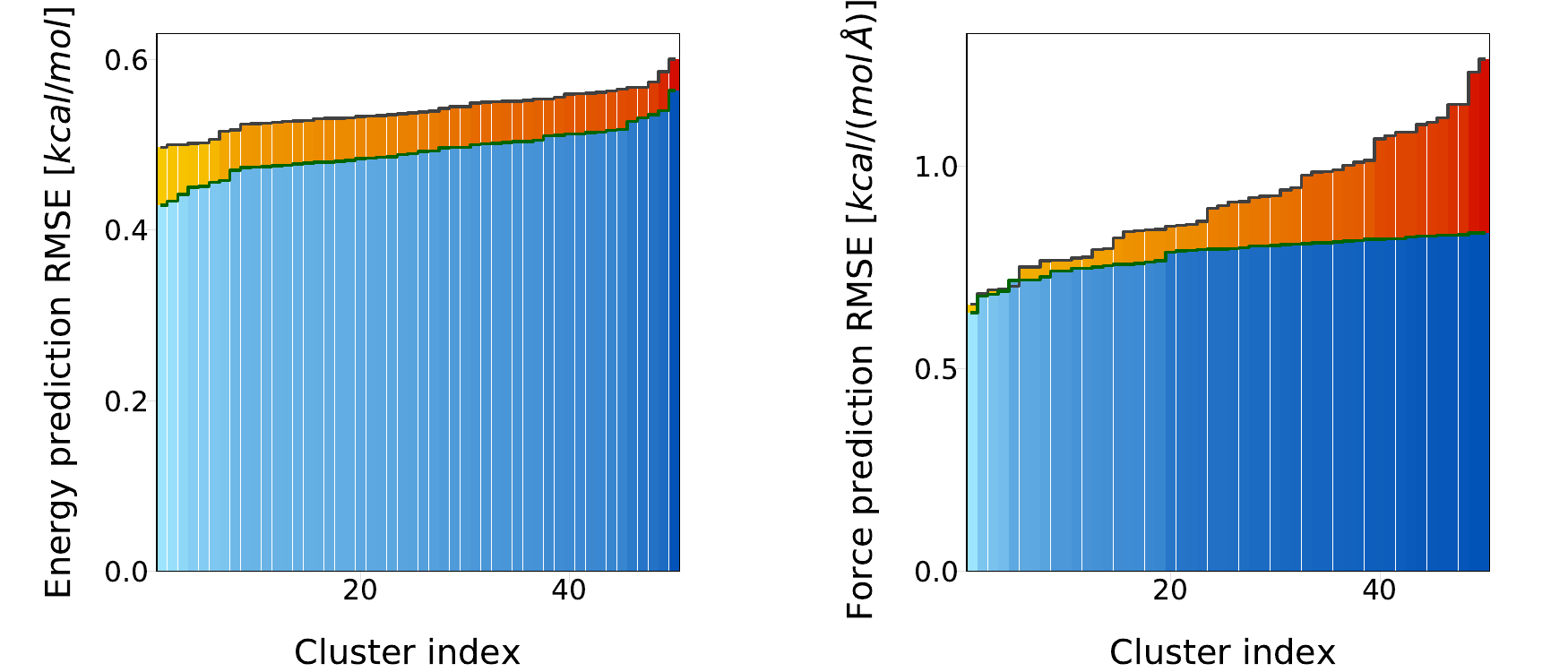}
    \caption{Energy (left) and force (right) prediction RMSE for a AcAla3NHMe SchNet default models on different clusters: default model (orange bars) compared to the improved model (blue bars). Each model consists of 6000 training points with identical training procedures and architecture.}
    \label{fig:AcAla3}
\end{figure*}

Our reference dataset was constructed via \textit{ab initio} molecular dynamics at 500~K with the FHI-aims software~\cite{FHIaims} wrapped with the i-PI package~\cite{IPI}, using the Perdew-Burke-Ernzerhof (PBE) exchange-correlation functional~\cite{MadeSimple} with tight settings and the Many-Body Dispersion (MBD) method~\cite{Ambrosetti2014, ATVDW} to account for van der Waals interactions. The dataset contains over 80k data points and covers at least three local minima. Our ML models were trained on 6k points, a number large enough to fully represent all physically different configurations given the intrinsic correlation between neighboring steps within MD trajectories. Figure~\ref{fig:AcAla3} shows the performance of two equal-size well-converged SchNet models trained using the default and improved training schemes for identical architectures. The details of the improved training procedure are similar to those in the \textit{Practical Application} section with two main differences: first, the number of training points added at each step is 500. Secondly, 150 fine clusters are created and several training points are extracted from each, where the number of points is weighted by the size of the fine cluster. 

Importantly, our method concentrates only on the improvement of the force predictions, where the RMSE of the default model for the worst clusters is about twice as large as that for the best one. The minor improvement in energy demonstrated in Figure~\ref{fig:AcAla3} is a nice accompanying effect which we were not aiming for. Although the overall improvement in RMSE for forces and energy drops only from $0.89$ to $0.8$~$kcal/(mol\,\AA)$ and $0.54$ to $0.49$~$kcal/mol$ respectively, the flattening of the errors for the force prediction can have noticeable results in practice. Improving the force predictions along with learning the energy within the SchNet model (or any other ML model) would usually require the employment of mixed loss functions, where the errors for energy and forces are minimized together. Such mixed loss functions are less efficient than pure ones, since optimizing two competing functions leads to sub-optimal results for each component. In contrast, the proposed method decouples the problem of improving the predictions for energy and forces. We use an energy-based loss function in the SchNet model while the forces are improved by manipulating the training set. As a result, we have a ML model with equally reliable energies and forces. Moreover, the errors in energy predictions also decrease, which would not happen if mixed loss functions were used. Alternatively, if our method instead focused on flattening the energy predictions, the improvements in the latter would be significantly more noticeable. However, this would come at the expense of force predictions, which would largely worsen as a result. 

Figure~\ref{fig:AcAla3} also reveals challenges for the developed method when applied to large systems. The cluster force predictions of the default model does not present a variance quite as large as the previously explored molecules. The main reason is that high-dimensional space (AcAla3NHMe has 42 atoms, i.e. 861 pairwise atomic distances) makes our clustering algorithms significantly weaker. Any distance metric loses meaning as the number of dimensions increases, and clustering algorithms rely on the latter to subdivide datasets. As a consequence, our clusters are ill-defined and contain larger overlaps between qualitatively different configurations. This reduces the resolution between well and poorly predicted parts of CS, decreasing the efficiency of the proposed method. We expect that reducing the size of descriptors by making use of dimensionality reduction techniques (such as kernel principal component analysis) would improve the efficiency of the clustering schemes and in turn make the developed approach reliable for systems containing hundreds and thousands of atoms.

\begin{figure*}[h!]
    \centering
    \includegraphics[width=1\textwidth]{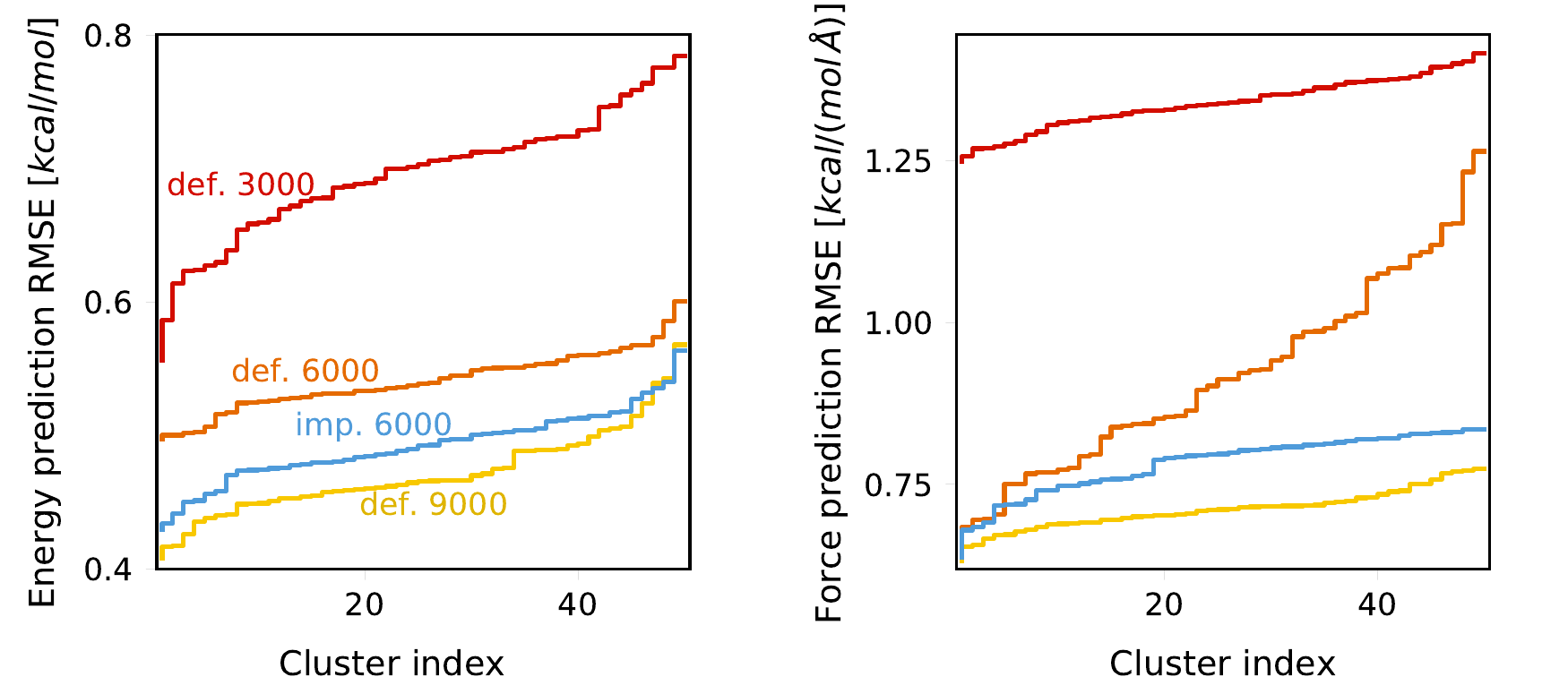}
    \caption{Energy (left) and force (right) prediction RMSE for a AcAla3NHMe SchNet default (orange) and improved (blue) models on different clusters. Comparing different size of training sets: 3000, 6000, 9000 for default and 6000 for improved.}
    \label{fig:AcAla3train}
\end{figure*}

Nevertheless, even without the aforementioned additional step, significant improvement can be found when applying our training methods to alanine tetrapeptide: both energy and force prediction errors are reduced for almost every cluster. Importantly, achieving the same improvement within the default training scheme would require adding more reference data to the training set. Figure~\ref{fig:AcAla3train} shows the energy and forces prediction accuracy of the SchNet models trained with different sizes of the training set. Depending on the size of the training set, one can observe three qualitatively different behaviors of the resulting SchNet models: 
(a) Whenever the training set contains insufficient data (the model with 3k points), the constructed MLFF demonstrates low accuracy across the entire CS for both energy and forces. In this limit, the force-based improved training method proposed in this work does little to improve the FF since the starting model cannot distinguish between poorly and well-predicted areas of the CS, 
(b) The training set contains enough data for the ML model to accurately learn the PES, but the forces are poorly predicted across CS (the model with 6k points), akin to previous examples (see the \textit{Improved Models} subsection). This is precisely the scenario for which the proposed improved training technique has been developed. By comparing the default and improved models with 6k training points, one can see a significant boost in accuracy for forces accompanied by a slight improvement of the PES reconstruction (here, an improved model of 6k points is comparable to a default model of 7.5-8k points). As such, the proposed training method gives an optimal compromise between data-efficiency and accuracy of ML models,
(c) Finally, a training set overloaded with reference data (the model with 9k points) leaves little room for improvement. Indeed in this case, the training contains all relevant configurations in the dataset (and by extension the validation set), such that the choice of training points becomes insignificant. 

\newpage

To compare the performance of the improved and default models, we ran constant-temperature MD simulations at 300~K and 400~K using the SchNet FF model. The time step was set to 0.5~fs to accurately reproduce fast hydrogen fluctuations in the molecule. Due to the size and high flexibility of the peptide, obtaining well-converged average energies requires MD trajectories of more than four million steps, equivalent to two nanoseconds. Simulations of this size come at prohibitively expensive computational costs for any accurate \textit{ab initio} method; MLFFs are the only way to perform them in practice. Note that our improved training procedure does come with higher computational costs (due to training the model multiple times), but the time spent on training is still very low compared to that of actually running the MD.

\begin{figure*}[h!]
    \centering
    \includegraphics[width=.5\textwidth]{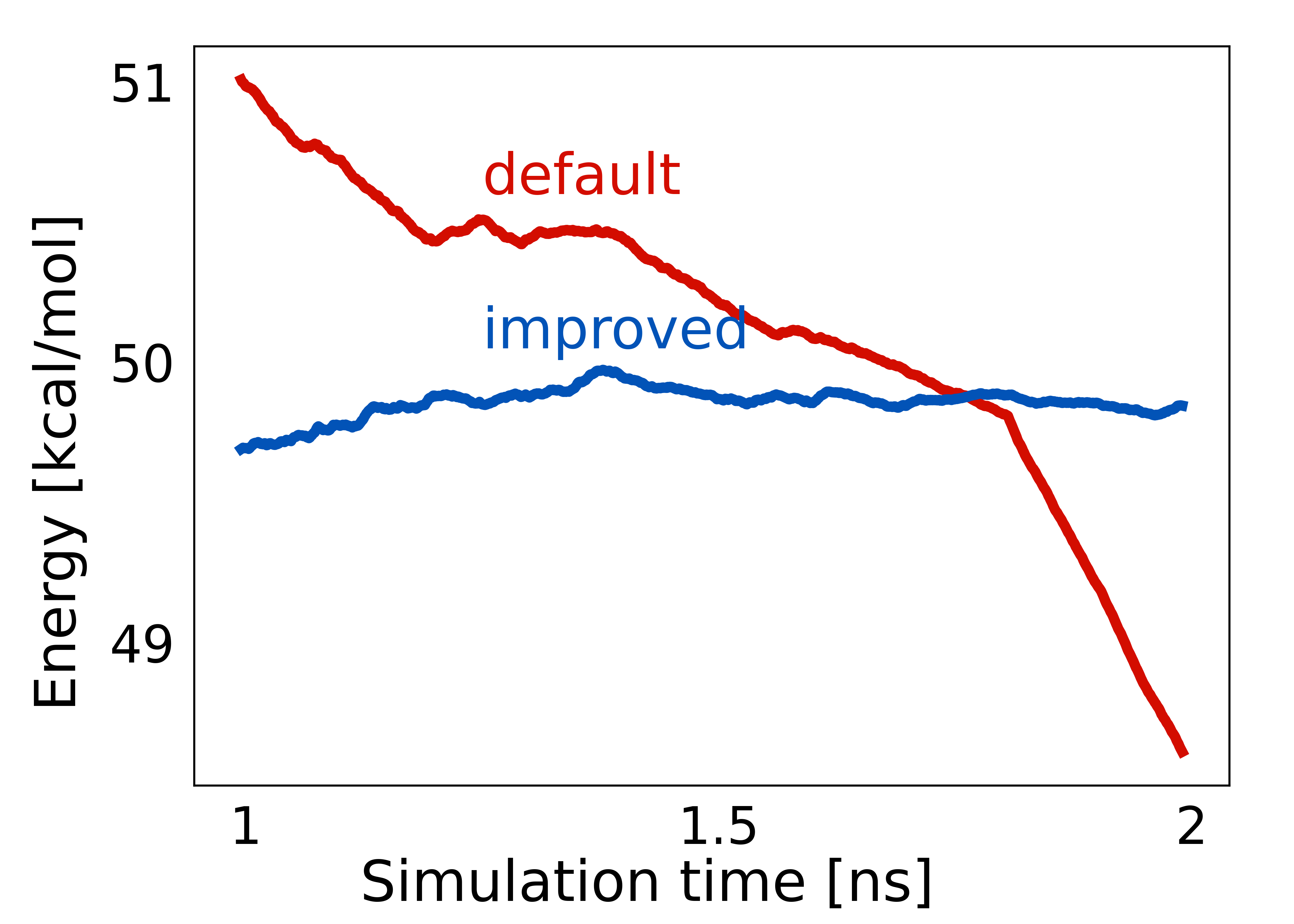}
    \caption{Average total energy as a function of simulation time for the default (red) and the improved (blue) SchNet models for the AcAla3NHMe molecule. The constant-temperature MD simulations have been done at 400~K with 0.5~fs time step.}
    \label{fig:AcAla3_400K}
\end{figure*}

At 300~K both models converge without any issues with a difference in average total energies of only 0.5~$kcal/mol$. The latter is within the accuracy of the ML models, see Figure~\ref{fig:AcAla3train}, meaning that both simulations give identical results. This is exactly what should be expected for a well trained ML model in its zone of comfort. At 400~K the situation changes drastically: the average total energy as a function of simulation time is shown in Figure~\ref{fig:AcAla3_400K}. One can see that while the improved model remains stable (as a zero energy level, we use the lowest potential energy in the reference dataset), the default one fails to reproduce the dynamics of the molecule at 400~K. The monotonic decay of the red curve advocates for the unreliability of the default training scheme for high-temperature simulations. As a result of wrong predictions, the molecule escapes the applicability range of the MLFF and we observe nonphysical results. Note that the training set was generated at 500~K, and thus contains all the information needed for 400~K MD simulations. 
Importantly, increasing the temperature to generate new reference data would require broader sampling of parts of CS with computationally expensive ~\textit{ab initio} methods --- an unacceptable scenario for growing molecule sizes.
Hence, the developed improved training scheme not only leads to quantitatively better predictions, but also qualitatively increases the applicability range of the ML models by boosting their reliability.

%%%%% DISCUSSION
\section{Conclusions} \label{Conclusions}
By leveraging supervised and unsupervised ML, we proposed a new strategy for improved training set selection for the construction of molecular machine learning force fields. We developed an automatic outlier detection method that exposed a noticeable bias in the predictive accuracy of the models towards common/in-equilibrium configurations at the expense of rarer/out-of-equilibrium ones, leading to entire regions of CS with significantly higher-than-average prediction errors. Our procedure is able to extract tiny subsets of molecular configurations representing nontrivial physical or chemical processes from an overwhelming amount of reference data. For example, a few hundred configurations with fingerprints of a shared proton in the salicylic acid molecule were found within 300k+ classical fluctuations around the equilibrium state.

The developed error analysis helped us optimise the training set choice, resulting in largely improved accuracy of ML models across all of CS---effectively ``flattening" the prediction error curve throughout input space. During the training process, we iteratively selected poorly predicted training points from different parts of CS to add to the training set. This ensured that it contained sufficient representation from every qualitatively different type of configurations in the reference dataset. Models born from this approach proved more reliable than those with training sets in-line with the dataset's inherent distributions and guarantee ``chemical accuracy" for the entire sampled CS. With the examples of small organic molecules and an alanine tetrapeptide we demonstrated that the developed training method leads to an optimal compromise between data-efficiency and accuracy of MLFFs, avoiding the need to generate extensive amounts of computationally expensive highly-accurate reference data for training sets. Along with quantitative reductions in prediction errors, the ML models trained on the optimised training sets offer qualitative improvements in reliability for practical applications. This is demonstrated on the example of high-temperature MD simulations for the alanine tetrapeptide. Future plans include combining the developed approach to dimensionality reduction techniques to extend the applicability range to systems consisting of hundreds and thousands of atoms.

While this paper focused on improving three specific ML models (GAP with SOAP descriptors and sGDML as kernel based approaches and SchNet as a neural network), all methods can easily be extended to any ML field presenting similar training problems. The code for the outlier detection and improved training is available in the open-source software MLFF on Github.~\cite{github}

\begin{acknowledgments}
We acknowledge financial support from the Luxembourg National Research (FNR) under the AFR project 14593813 and C19/MS/13718694/QML-FLEX, FNR DTU-PRIDE MASSENA, and the European Research Council (ERC-CoG grant BeStMo).
\end{acknowledgments}

\section{Data availability}
The data that support the findings of this study are openly available on the sGDML \cite{gdml, sgdml} official website and in the supplementary material.

\bibliography{bib}

\end{document}